\documentclass[fleqn,usenatbib]{mnras}

\usepackage{newtxtext,newtxmath}
\usepackage{longtable}
\usepackage{array}
\usepackage[T1]{fontenc}

\DeclareRobustCommand{\VAN}[3]{#2}
\let\VANthebibliography\thebibliography
\def\thebibliography{\DeclareRobustCommand{\VAN}[3]{##3}\VANthebibliography}

\usepackage{graphicx}	
\usepackage{amsmath}	
\usepackage{booktabs} 
\usepackage{natbib}  
\usepackage{xcolor}
\usepackage{orcidlink}

\title[Constraining Lens Galaxy Mass Evolution with Lensed Quasars]{New Dynamical Measurements from a Lensed Quasar Sample: Joint Analysis Constrains the Mass Profile Evolution of Lens Galaxies}

\author[Z. Guo, et al.]{
Ziyu Guo\orcidlink{0009-0002-0711-2948}, \textsuperscript{1}
   Yun Chen\orcidlink{0000-0001-8919-7409},\textsuperscript{2,3}\thanks{Email: chenyun@bao.ac.cn} 
       Yiping Shu,\textsuperscript{4}
    \thanks{Email:yiping.shu@pmo.ac.cn}
    Jiaze Gao, \textsuperscript{5}
    Hui Li,\textsuperscript{6}
     Zizhao He,\textsuperscript{7,8,4}
      and Jun Wang
      \textsuperscript{1}\thanks{Email: wjun@ynu.edu.cn}
\\
   $^1$ School of Physics and Astronomy, Yunnan University, Kunming 650091, China;\\
    $^2$ National Astronomical Observatories, Chinese Academy of Sciences, No. 20A Datun Road, Beijing 100101, China;\\
   $^3$ School of Astronomy and Space Science, University of Chinese Academy of Sciences, Beijing 100049, China;\\
   $^4$ Purple Mountain Observatory, Chinese Academy of Sciences, No. 10 Yuan Hua Road, Nanjing 210023, China\\
   $^5$ Institute of Theoretical Physics, School of Physics, Dalian University of Technology, Dalian 116024, China;\\
   $^6$ Institute for Astrophysics, School of Physics, Zhengzhou University, Zhengzhou, 450001, China;\\
   $^7$ Department of Physics, Nanchang University, Nanchang, 330031, China\\
   $^8$ Center for Relativistic Astrophysics and High Energy Physics, Nanchang University, Nanchang, 330031, China\\
}

\date{Accepted XXX. Received YYY; in original form ZZZ}

\pubyear{\the\year{}}

\begin{document}
\label{firstpage}
\pagerange{\pageref{firstpage}--\pageref{lastpage}}
\maketitle

\begin{abstract}

We present a systematic study of the internal mass structure of early-type galaxies (ETGs) based on 106 galaxy-scale strong gravitational lenses with background quasars, all having spectroscopic redshifts. 
From this parent sample, we select 24 systems that possess the high-quality ancillary data necessary for a joint analysis of strong-lensing geometry and stellar kinematics, with lens redshifts spanning $0.195 \leq z_l \leq 0.867$.
A key contribution is the derivation of new single-aperture stellar velocity dispersions for 11 lens galaxies via an iterative spectroscopic fitting procedure that mitigates quasar contamination, providing previously unavailable data. We model the total mass-density profile as a power law, $\rho \propto r^{-\gamma}$,
and parameterise its logarithmic slope as $\gamma = \gamma_0 + \gamma_z \cdot z_l + \gamma_s \cdot \log \tilde{\Sigma}$, 
where $z_l$ is the lens redshift and $\tilde{\Sigma}$ the surface mass density. Within a flat $\Lambda$CDM framework and using DESI BAO measurements as a prior, we constrain the parameters via Monte Carlo nested sampling to $\gamma_0 = 1.62^{+0.11}_{-0.12}$, $\gamma_z = -0.35^{+0.08}_{-0.09}$, and $\gamma_s = 0.37^{+0.08}_{-0.07}$ ($68\%$ confidence intervals). Our results robustly demonstrate that $\gamma$ increases with surface mass density ($\gamma_s > 0$) and decreases with redshift ($\gamma_z < 0$). This implies that, at fixed redshift, galaxies with denser stellar cores have steeper mass profiles, while at fixed density, profiles become shallower at higher redshifts. By successfully applying the joint lensing--dynamics method to a substantial, independently acquired sample of lensed quasars, this work provides crucial validation of structural trends previously observed in galaxy--galaxy lensing systems, reinforcing the established evolutionary picture for massive ETGs and establishing lensed quasars as a potent probe of galaxy structure.

\end{abstract}


\section{Introduction}
\label{sect:intro}

The internal mass structure of massive ETGs serves as a fossil record of their assembly history. Within the hierarchical framework of $\Lambda$CDM cosmology, ETG formation is driven by a combination of dissipative processes, mergers, and feedback, each of which imprints distinct signatures on the radial mass density profiles, $\rho(r)$ \citep{Remus2013, Derkenne2021}. Constraining the shape and evolution of these profiles is therefore critical for testing galaxy formation models. Strong gravitational lensing provides a unique and powerful tool for this purpose, as it directly measures the projected total mass distribution independently of the dynamical state or nature of the matter \citep{Bolton2006, Shu_2016, Li2018, Sonnenfeld2025}.

Whilst galaxy-galaxy strong lensing (GGSL) systems have been extensively used to study the average mass profiles of ETGs and their evolution \citep{Auger2010, Sonnenfeld2013b, Chen2019, 2025arXiv251108030W}, the rapidly growing sample of galaxy-scale lenses with background quasars -- galaxy-quasar strong lensing (GQSL) -- offers a complementary and independent avenue. Lensed quasars present distinct observational characteristics: their point-like images enable extremely precise astrometry for measuring Einstein radii, and their broad emission lines typically yield secure source redshifts. These advantages are balanced by specific challenges, most notably the contamination of the lens galaxy spectrum by the bright background quasar, which complicates the measurement of stellar kinematics -- a key component for joint lensing and dynamical analysis.
\citet{Shajib_2026} introduced a spectral decomposition method specifically designed to mitigate such contamination, representing an important recent advance. This approach enables explicit modelling and subtraction of quasar contamination in strong lensing data. However, because it typically relies on integral field unit (IFU) observations, its direct application to large samples of strong lensing systems is limited.

Previous joint lensing and dynamics studies, primarily using GGSL systems, have established that the total mass density profile of massive ETGs within the effective radius is well approximated by a power law, $\rho \propto r^{-\gamma}$, with a mean slope near isothermal ($\gamma \sim 2$) \citep{Koopmans2006, Grillo2008}. More detailed analyses have revealed subtle yet significant trends, indicating that $\gamma$ increases with stellar mass or central surface density \citep{Auger2010} and exhibits mild evolution with redshift \citep{Ruff2011, Bolton2012, Sonnenfeld2013a}. The most comprehensive studies by \citet{Sonnenfeld2013a} and \citet{Chen2019} have quantitatively parameterised these dependencies, finding $\gamma$ to correlate positively with surface density and negatively with redshift. These trends are consistent with a picture of inside-out growth, in which higher-density, later-forming galaxies possess more centrally concentrated mass distributions.

Extending this gravitational-dynamical method to lensed quasars is a logical and necessary progression. It not only provides an independent verification of results derived from GGSL systems, potentially probing different ranges of lens redshift and mass, but also validates the methodology for a class of lenses crucial for time-delay cosmography \citep{Treu2016, TDCOSMO2025}, where a precise understanding of the lens mass profile is essential for accurate Hubble constant measurements. However, such an extension has been limited by the availability of stellar velocity dispersions for the lens galaxies, which are difficult to measure against the glare of the quasar.

The primary aim of this work is to construct and exploit a substantial, homogeneously analysed sample of GQSL systems to overcome the longstanding limitation in stellar kinematics measurement. By acquiring new spectroscopic data and implementing a robust analysis pipeline to mitigate quasar contamination, we perform the first systematic joint lensing and dynamical study on such a sample. This dataset serves as a critical new resource for independently testing galaxy evolution models.

This study robustly demonstrates that GQSL sample is a powerful and reliable avenue for constraining galactic mass profiles. By successfully applying the joint lensing–dynamics method to a substantial, independently acquired sample, we validate this population as a crucial and complementary probe. It thereby opens a new, independent window into the structural evolution of massive galaxies, one that is particularly potent for probing higher-redshift regimes and strengthening the foundations of time-delay cosmography.

The paper is structured as follows:  Section~\ref{sect:data} details the construction of our lensed quasar sample and presents the newly acquired kinematic and photometric measurements. Section~\ref{sect:method} outlines the formalism for the joint lensing and dynamics analysis. Section~\ref{sect:analysis} presents the resultant constraints on the mass-profile parameters and provides some discussion. Finally, Section~\ref{sect:conclusion} synthesises our principal findings and discusses their implications within the broader context of galaxy evolution, alongside prospects for future work.

\section{Data Sample}
\label{sect:data}

\subsection{Sample construction and selection criteria}
\label{sect:sample}

A well-defined and homogeneous sample of galaxy-scale strong gravitational lens systems is essential for the joint lensing and stellar dynamics methodology detailed in Section~\ref{sect:method}. As our focus is on the evolution of ETGs, we selected systems that are isolated, with no significant substructure or massive nearby companions that could perturb the gravitational potential.

The analysis requires precise measurements of the following key observables for each system: (i) the lens and source redshifts ($z_l$ and $z_s$); (ii) the Einstein radius ($\theta_E$) and the effective (half-light) radius of the lens galaxy ($\theta_{\mathrm{eff}}$), both derived from high-resolution imaging; and (iii) the central stellar velocity dispersion ($\sigma_{\mathrm{ap}}$) measured within a well-defined spectroscopic aperture of angular radius $\theta_{\mathrm{ap}}$.

\begin{figure*}
    \centering
    \includegraphics[width=1.0\linewidth]{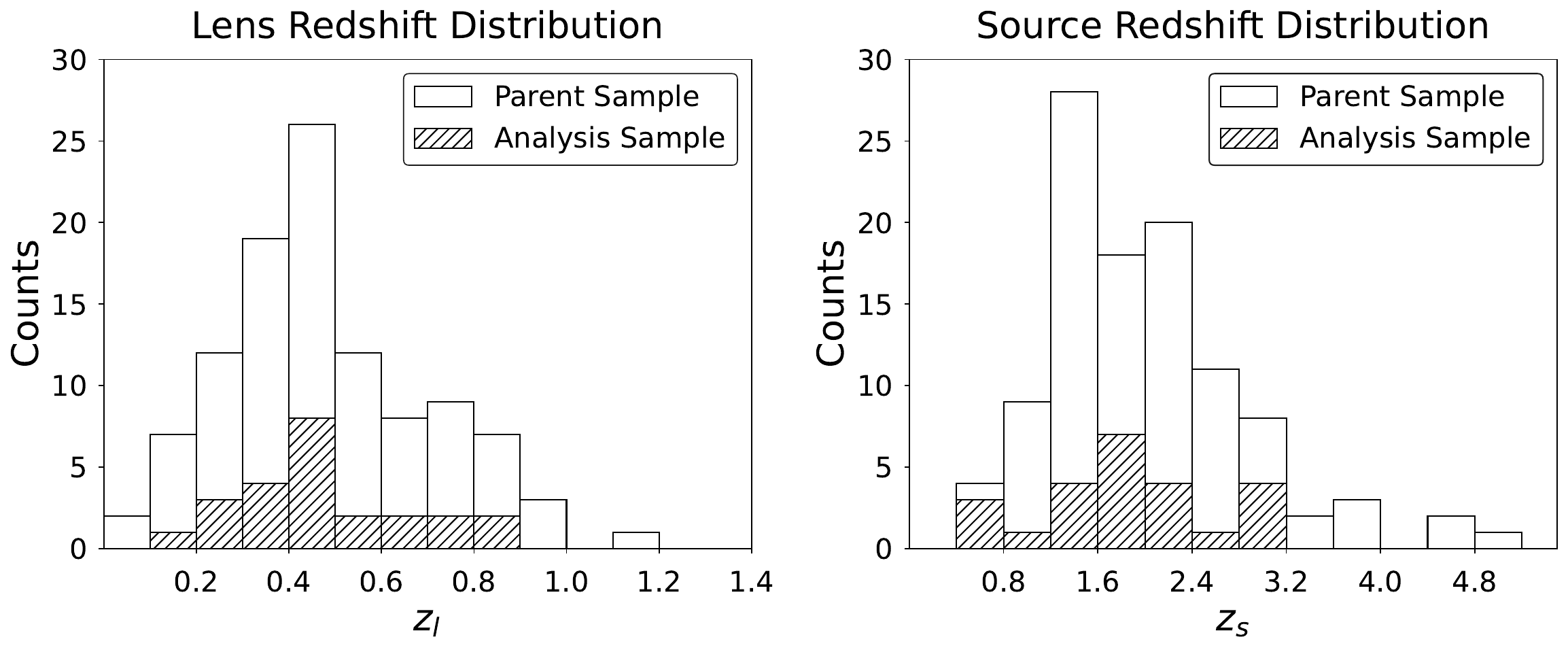}
    \caption{Redshift distributions of the lens and source samples. \textbf{Left panel:} distribution of lens galaxy redshifts ($z_l$). \textbf{Right panel:} distribution of background source (quasar) redshifts ($z_s$). The black bars represent the Parent Sample of 106 systems, and the black shaded region represents the Analysis Sample of 24 systems used for the joint analysis.}
    \label{fig:all_samples_redshift_distribution}
\end{figure*}

We initiated our sample compilation using the Gravitationally Lensed Quasar Database\footnote{\url{https://research.ast.cam.ac.uk/lensedquasars/index.html}}, which offers a comprehensive collection of lens systems reported in the literature up to 2019. We then augmented this catalogue with newly discovered systems reported after 2019 and updated the parameters of previously known lenses by incorporating revised measurements from subsequent studies wherever possible, prioritising the most recent and precise results. Through this iterative process of compilation and revision, we assembled a comprehensive sample of all currently known galaxy-scale strong gravitational lens systems with quasars as background sources, totalling 344 systems. From this, we selected systems with spectroscopically confirmed $z_l$, $z_s$, and $\theta_{\mathrm{max}}$ for ETG lenses, yielding a parent sample of 106 systems. Among these, only 13 systems simultaneously have reported values of $\theta_{\mathrm{E}}$, $\theta_{\mathrm{eff}}$, and $\sigma_{\mathrm{obs}}$, and are therefore directly usable for the analysis presented in this work. To expand the analysis sample, we measured $\sigma_{\mathrm{obs}}$ for 11 systems and $\theta_{\mathrm{eff}}$ for 4 systems using publicly available spectroscopic and imaging data. This increased the number of systems in the analysis sample from 13 to 24, thereby enabling the statistical analysis. The lens redshifts in the final analysis sample span $0.195 \leq z_l \leq 0.867$, whilst the source redshifts range from $0.654$ to $3.140$. The redshift distributions for both the parent and analysis samples are shown in Figure~\ref{fig:all_samples_redshift_distribution}.

The complete analysis sample, including all relevant measurements and their references, is listed in Table~\ref{tab:lens_data}. This table presents the fundamental data for our study. The effective radius ($\theta_{\mathrm{eff}}$) and the stellar velocity dispersion ($\sigma_{\mathrm{obs}}$) are the two critical inputs for the joint analysis that were not available in the literature for all systems. Values marked with an asterisk ($^*$) in these columns indicate new measurements obtained in this work, comprising velocity dispersions for 11 systems and effective radii for 4 systems. The following subsections detail how we acquired these missing data.

\begin{table*}
\tabcolsep 0.2cm
    \centering
    \caption{Observational properties of the 24 lensed quasar systems used for the joint lensing and dynamics analysis.}
    \begin{tabular}{lcccccccl}
        \hline\noalign{\smallskip}
        Name &  $z_l$ & $z_s$ & $\theta_{\text{E}}$  & $\theta_{\text{eff}}$  & $\theta_{\text{ap}}$  & $\sigma_{\text{obs}}$  & Reference \\
         &  &  &  (arcsec) & (arcsec) & (arcsec) & (km/s) & \\
        \hline\noalign{\smallskip}
        SDSSJ0114+0722 & 0.408 & 1.828 & 0.83  & 0.84  & 0.75  & 222.31$\pm$20$^*$ & \citet{More2016}\\
        PSJ0147+4630  &  0.678 & 2.341 & 1.90  & 3.45  & 2.59 & 283$\pm$32.31 &\citet{shajib2019, Mozumdar2023}\\
        J0203+1612    &  0.488 & 2.18  & 1.36  & 3.15  & 1.00 & 206.88$\pm$50$^*$ &\citet{Lemon2019}\\
        HE0435-1223   &  0.454 & 1.689 & 1.22  & 1.8   & 0.318 & 226.6$\pm$5.8 &\citet{TDCOSMO2025}\\
        J0659+1629    &  0.766 & 3.09  & 2.35  & 0.41  & 7.65  & 326$\pm$30.87 &\citet{Ertl2023, Mozumdar2023}\\
        J0818-2613    &  0.866 & 2.164 & 2.896 & 2.43  & 2.59  & 392$\pm$50.99 &\citet{Schmidt2023, Mozumdar2023}\\
        SDSSJ0819+5356 &  0.294 & 2.237 & 2.057 & 5.84  & 1.00 & 301.07$\pm$11$^*$ &\citet{Inada2009}\\
        HSCJ0918-0220 &  0.460  & 0.803 & 1.28  & 0.40  & 0.75   & 360.48$\pm$30$^*$ &\citet{Jaelani2021}\\
        SDSSJ0924+0219 &  0.393 & 1.524 & 0.95 & 0.436 & 2.59  & 209$\pm$15.00 &\citet{Chen2022, Mozumdar2023}\\
        J0941+0518    &  0.34  & 1.54  & 2.72  & 2.29$^*$  & 1.00 & 311.10$\pm$12$^*$ & \citet{Lemon2018}\\
        J0949+4208    &  0.51  & 1.27  & 1.23  & 2.22  & 0.75  & 290.44$\pm$50$^*$ &\citet{Lemon2018}\\
        COSMOS5921+0638 &  0.551 & 3.14  & 0.70  & 0.49  & 0.75  & 186.42$\pm$30$^*$ &\citet{Faure2008}\\
        PG1115+080    &  0.311 & 1.722 & 1.08  & 0.45  & 0.318 & 235.7$\pm$6.6&\citet{TDCOSMO2025}\\
        RXJ1131-1231  &  0.295 & 0.654 & 1.63  & 1.91  & 0.318 & 303$\pm$8.3 &\citet{TDCOSMO2025}\\
        SDSSJ1206+4332 &  0.745 & 1.789 & 1.00  & 0.29  & 0.318 & 290.5$\pm$9.5 &\citet{TDCOSMO2025}\\
        HSCJ1215-0058  &  0.45912 & 2.8817 & 2.342 & 3.015$^*$ & 0.75  & 219.72$\pm$60$^*$ &\citet{He2025}\\
        HSCJ1220+0112  &  0.48755 & 1.7081 & 0.9584 & 0.729$^*$ & 0.75  & 222.03$\pm$40$^*$ &\citet{He2025}\\
        HST1411+5211  & 0.465 & 2.811 & 0.615 & 0.61  & 0.9  & 174$\pm$20 &\citet{Lubin2000, Hamana2005}\\
        SDSSJ1433+6007 & 0.407 & 2.737 & 1.71  & 1.10  & 2.59 & 261$\pm$9.22 &\citet{shajib2019, Mozumdar2023}\\
        SDSSJ1524+4409 & 0.320  & 1.210  & 0.79  & 0.55$^*$  & 0.75  & 222.26$\pm$15$^*$ &\citet{Oguri2008}\\
        SDSSJ1640+1932 & 0.195 & 0.778 & 2.49  & 2.15  & 1.00 & 372.85$\pm$8$^*$ &\citet{Wang2017}\\
        WFI2033-4723  & 0.661 & 1.66 & 0.94  & 1.97  & 0.318 & 210.7$\pm$10.5 &\citet{TDCOSMO2025}\\
        WGD2038-4008  & 0.230 & 0.777 & 1.38  & 2.223 & 1.5  & 254.7$\pm$16.3 &\citet{TDCOSMO2025}\\
        B2045+265     & 0.8673 & 1.28  & 0.638 & 0.411 & 1.2  & 213$\pm$23 &\citet{Mckean2007, Hamana2005}\\
        \hline
    \end{tabular}
    \label{tab:lens_data}
    \vspace{2mm}

    \small
    \textbf{Notes.} Columns: (1) Lens system name; (2) Lens galaxy redshift; (3) Source quasar redshift; (4) Einstein radius in arcseconds; (5) Effective radius of the lens galaxy in arcseconds; values marked with an asterisk ($^*$) are new estimates from this work (Section~\ref{sect:reff}); (6) Angular radius of the spectroscopic aperture in arcseconds, for slit observations, the aperture correction formula has been applied; (7) Observed velocity dispersion (km s$^{-1}$) measured within the corresponding aperture; values marked with an asterisk ($^*$) are new measurements from this work (Section~\ref{sect:spec}); (8) References for columns (4) to (7). The asterisks denote new measurements from this work: velocity dispersions for 11 systems and effective radii for 4 systems.\\
\end{table*}

\subsection{Data acquisition and spectroscopic measurements}
\label{sect:spec}

As noted in Section~\ref{sect:sample}, stellar velocity dispersions were already available in the literature for 13 systems in our parent sample. To increase the sample size and ensure measurement homogeneity, we sought to obtain dispersions for the remaining systems. 
We performed a cross-match between known GQSL systems lacking stellar velocity dispersion measurements and the Baryon Oscillation Spectroscopic Survey (BOSS; \citealt{Dawson2013}) of data release DR16 of the Sloan Digital Sky Survey-III (SDSS-III), as well as the first data release (DR1) of the Dark Energy Spectroscopic Instrument (DESI; \citealt{DESI2025_DR1}), using a matching radius of 0.7 arcsec. 
This yielded 70 matches. For each system we selected the spectrum with the highest signal-to-noise ratio and visually inspected it; three spectra lacking reliable absorption features were discarded. This effort produced new velocity dispersion measurements for 11 systems: four derived from SDSS-III BOSS spectra and seven from DESI DR1 spectra.

Stellar velocity dispersions were measured using the Penalised Pixel-Fitting (\texttt{pPXF}) software package \citep{Cappellari2004, Cappellari2017, Cappellari2023}. We employed the UV-extended E-MILES simple stellar population models \citep{Vazdekis2016} as stellar templates. A major challenge is the substantial contamination of the lens spectrum by emission lines from the blended quasar images, which necessitates masking of the affected wavelength regions. To address this, we implemented an iterative masking procedure:
(1) perform an initial \texttt{pPXF} fit to the spectrum;
(2) using the known source redshift $z_s$, compute the expected positions of the quasar emission lines and evaluate the mean of the normalised residuals within a $\pm20$\AA\ window. For any region where the mean normalised residual exceeds 0.7, we mask the interval and expand the mask outward in steps of 10\AA\ until the residual falls below the threshold; all other regions remain unmasked.
The initial $\pm20$\AA\ window was adopted as a physically motivated seed scale, based on the characteristic widths of quasar broad emission lines. For the quasars in our analysis sample, which have a mean source redshift of $z_s = 1.83$, typical broad-line widths correspond to observed-frame scales of several tens of \AA. The residual threshold of 0.7 and the 10\AA\ expansion step were chosen empirically from tests on representative spectra to balance effective contamination removal against unnecessary masking of the adjacent continuum;
(3) apply the mask and repeat the fit.

\begin{figure*}
    \centering
    \includegraphics[width=.98\linewidth]{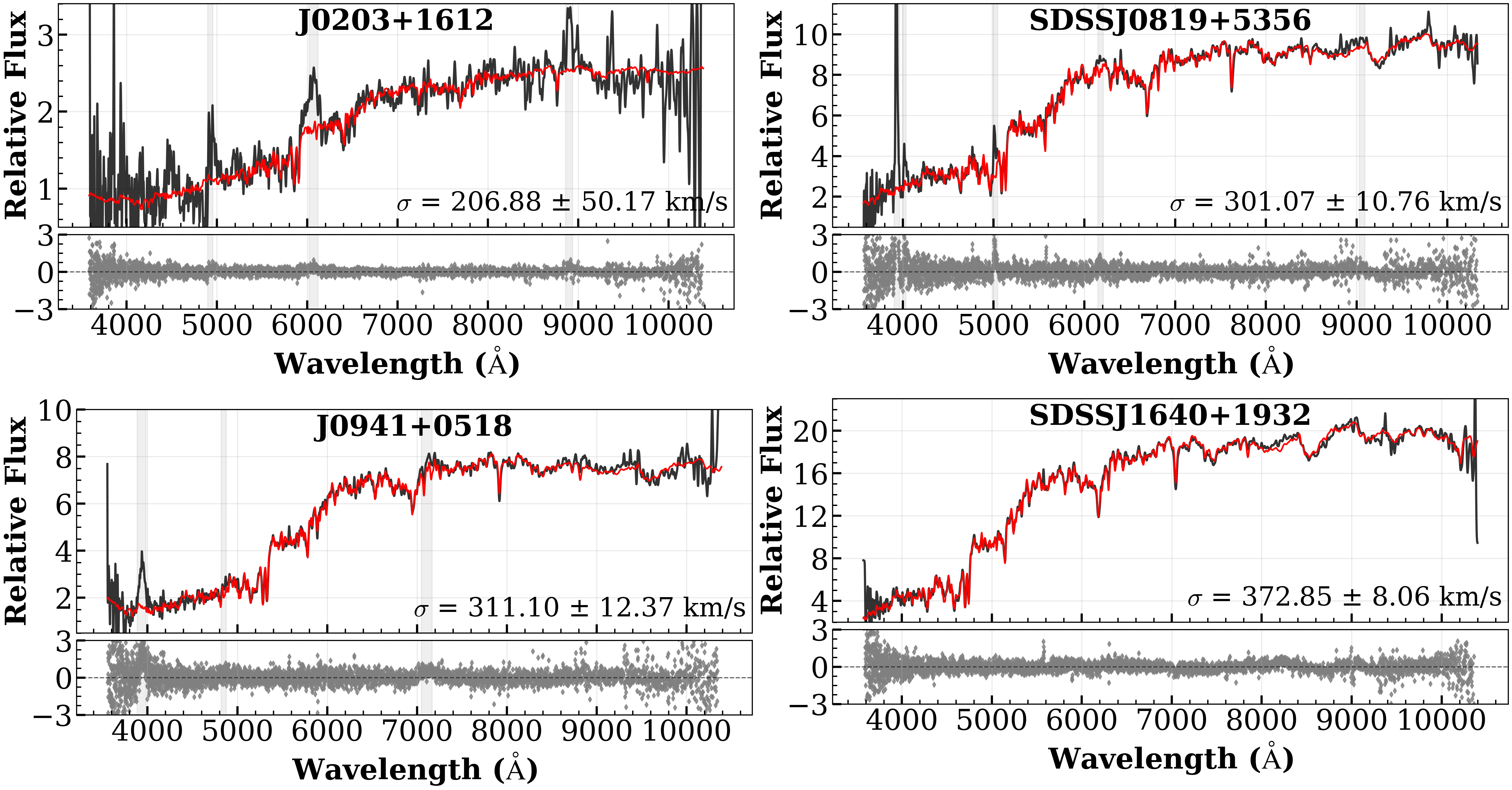}
    \caption{Stellar velocity dispersion measurements for the four lensing galaxies, derived from SDSS-III BOSS spectra. Each system is displayed in two panels. \textbf{Upper panel:} The black line shows the extracted and smoothed one-dimensional spectrum of the galaxy, with the wavelength range adjusted to emphasise absorption features. The red line indicates the best-fitting model obtained from the \texttt{pPXF} software applied to the unsmoothed spectrum. Grey bands mark regions masked during the fitting process owing to contamination by quasar emission lines. \textbf{Lower panel:} The normalised residuals.}
    \label{velocity_dispersion_from_BOSS}
\end{figure*}

\begin{figure*}
    \centering
    \includegraphics[width=.98\linewidth]{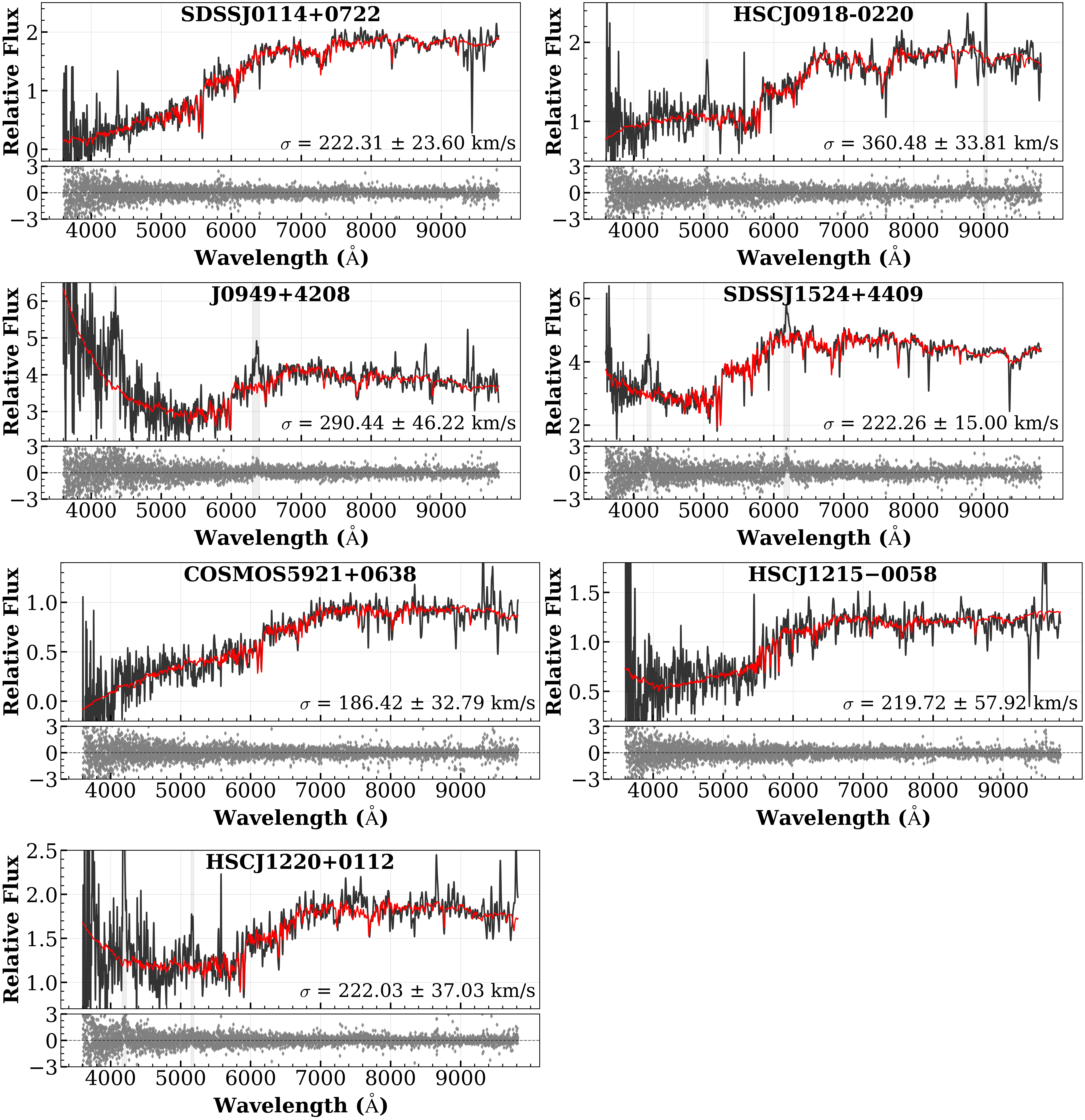}
    \caption{Stellar velocity dispersion measurements for the seven lensing galaxies, derived from DESI DR1 spectra. The layout follows that of Figure~\ref{velocity_dispersion_from_BOSS}, showing the results of \texttt{pPXF} fitting to the DESI spectral data for velocity dispersion.}
    \label{velocity_dispersion_from_DESI}
\end{figure*}

Examination of the velocity dispersion estimates obtained through this procedure shows that the normalised residuals outside the masked emission-line regions remain within $\pm1$ and exhibit no significant structure, confirming the robustness of our measurements. Moreover, the BOSS pipeline provides an independently measured velocity dispersion for each galaxy, derived using different stellar templates and fitting procedures.
In contrast to the standard BOSS pipeline, our procedure explicitly handles quasar‑contaminated wavelength regions using an iterative masking scheme, specifically designed to mitigate a key source of systematic error in lens‑galaxy spectroscopy. For the four systems in our sample that have official BOSS measurements, our velocity dispersion estimates agree with the BOSS values to within $1\sigma$, whilst the associated uncertainties are reduced by an average of about 25\% – with individual improvements ranging from 20\% to 31\%.
The fitting results for the BOSS and DESI spectra are presented in Figures~\ref{velocity_dispersion_from_BOSS} and~\ref{velocity_dispersion_from_DESI}, respectively.

\subsection{Measurement of effective radii}
\label{sect:reff}

For systems lacking published measurements of the effective radius ($\theta_{\mathrm{eff}}$), we performed our own photometric modelling using $z$-band imaging from the DESI Legacy Imaging Surveys (DESI-LS; \citealt{Dey2019}). The two-dimensional surface brightness distribution of the lens galaxy was modelled using the \texttt{lenstronomy} software package \citep{Birrer2018,Birrer2021}.

\begin{figure*}
    \centering
    \includegraphics[width=1\linewidth]{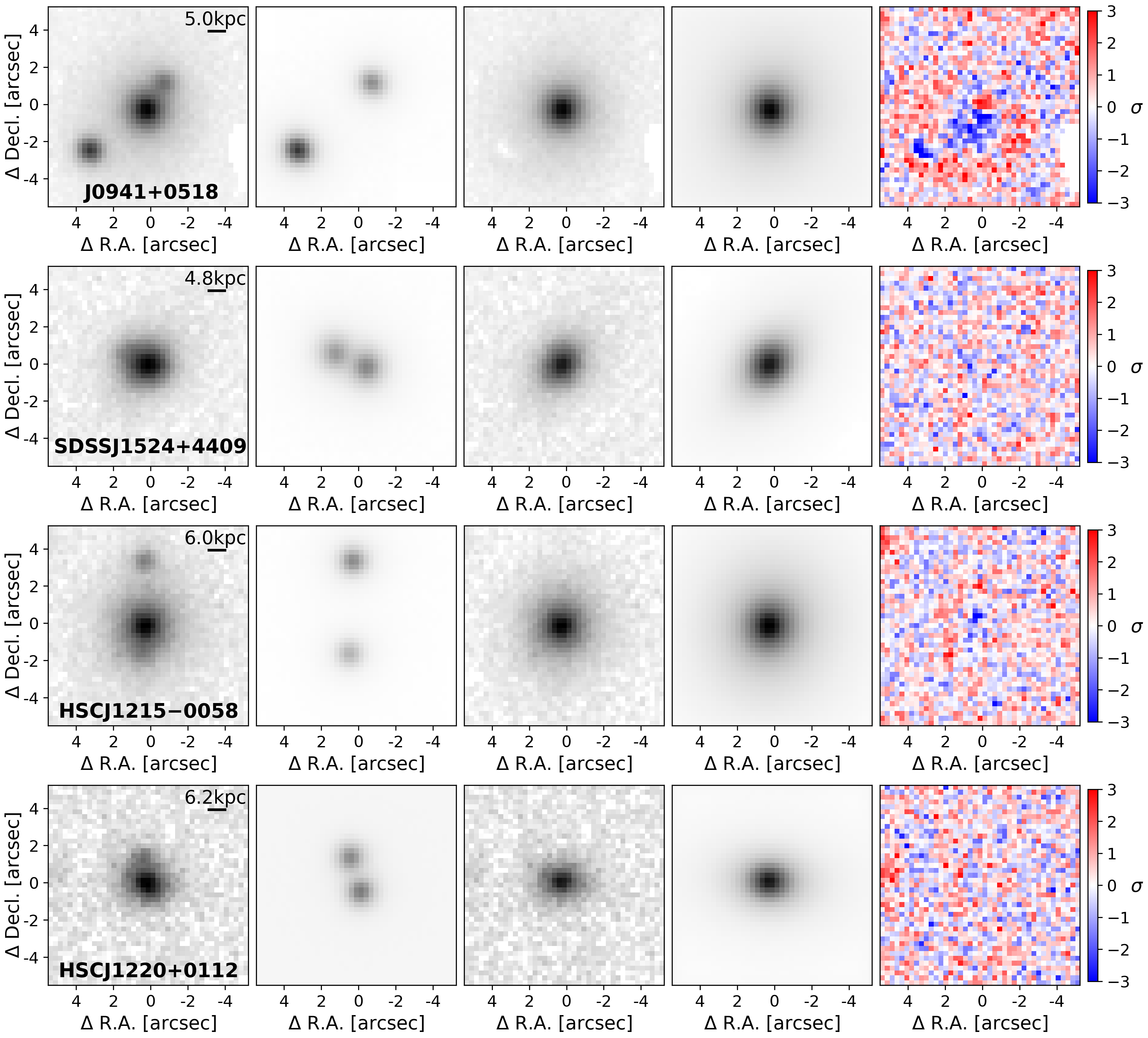}
    \caption{Half-light radius fitting results. Each row corresponds to a lens system. Columns show (from left to right): \textbf{(1)} the original DESI-LS image; \textbf{(2)} the model for the lensed quasar images; \textbf{(3)} the quasar-light-subtracted data used for fitting the galaxy light;  \textbf{(4)} the best-fitting de Vaucouleurs galaxy light model; and \textbf{(5)} the normalised residual map.}
    \label{effective_radius}
\end{figure*}

The surface brightness distribution of the lens galaxy was modelled using a composite surface-brightness model comprising a de Vaucouleurs component for the lens galaxy and point-source components for the lensed images of the background quasar. The de Vaucouleurs profile, a special case of the S\'{e}rsic model with index $n = 4$, is given by
\begin{equation}
    I(R) = I_e \exp\left\{ -b_4\left[ \left( \frac{R}{R_e} \right)^{1/4} - 1 \right] \right\},
\end{equation}
where $I_e$ is the surface brightness at the effective radius $R_e$, and $b_4 = 7.669$. The multiple quasar images were modelled as point sources using the provided point spread function (PSF).
Consequently, only two parameters were required to fit each image: the centroid coordinates and the amplitude. A prior distribution was adopted for the centroid coordinates of each quasar image, centred on the brightest pixel with an allowable deviation of $\pm1$ pixel. For the lens galaxy centre coordinates, a prior centred on the brightest pixel was employed, permitting a deviation of up to $\pm2$ pixels.
With the exception of J0941+0518, the normalised residuals of the final fits are predominantly within $\pm3$ and exhibit no significant systematic structure, indicating that our fitting procedure is highly robust. In the case of J0941+0518, however, residual structures are evident, which we attribute to possible lensing of the quasar host galaxy. Detailed results are presented in Figure~\ref{effective_radius}.
In Table~\ref{tab:lens_data}, effective radius values marked with an asterisk ($^*$) indicate measurements obtained through this work.

\section{Methodology}
\label{sect:method}
\subsection{The joint lensing and dynamics framework}

This study employs the gravitational-dynamical mass combination method to constrain the internal mass structure of early-type lens galaxies. The method rests on the fundamental equivalence, within the framework of General Relativity, between the gravitational mass derived from strong lensing geometry and the dynamical mass inferred from stellar kinematics, both measured within the Einstein radius:
\begin{equation} \label{eq:Mgrl_Mdyn}
M_{\mathrm{grl}}^E = M_{\mathrm{dyn}}^E.
\end{equation}
Here, $M_{\mathrm{grl}}^E$ denotes the projected mass within the Einstein radius inferred from lensing. On the dynamical side, the three‑dimensional mass model must be projected to obtain the corresponding two‑dimensional mass within the same radius, denoted $M_{\mathrm{dyn}}^E$, thereby enabling a direct comparison with the lensing measurement.
By jointly modeling these two independent mass estimates, we can constrain the parameters governing the galaxy's mass-density profile, most notably its power-law slope $\gamma$.

\subsection{Gravitational mass from strong lensing}

Under the assumption of a spherically symmetric lens, the gravitational mass enclosed within the Einstein radius $\theta_E$ is directly determined by the lens geometry and cosmological distances:
\begin{equation} \label{eq:Mgrl}
M_{\mathrm{grl}}^E = \frac{c^2}{4G}\frac{D_l D_s}{D_{ls}}\theta_E^2,
\end{equation}
where $c$ is the speed of light, $G$ the gravitational constant, and $D_{l}$, $D_{s}$, and $D_{ls}$ are the angular diameter distances to the lens, to the source, and between the lens and source, respectively. This expression provides a purely geometric mass estimate, independent of the dynamical state or the nature of the matter.

\subsection{Dynamical mass modeling and velocity dispersion prediction}

To relate the dynamical mass $M_{\mathrm{dyn}}^E$ to the observable stellar velocity dispersion, we must adopt a parameterized model for the lens galaxy. We follow the standard approach in lensing dynamics \citep{2006EAS....20..161K, Chen2019} and make the following key assumptions:
\begin{enumerate}
    \item The system is in dynamical equilibrium and spherically symmetric.
    \item The total mass-density profile $\rho(r)$ and the stellar luminosity density profile $\nu(r)$ follow power laws:
    \begin{equation} \label{eq:rho}
        \rho(r) = \rho_0 \left( \frac{r}{r_0} \right)^{-\gamma}, \quad
        \nu(r) = \nu_0 \left( \frac{r}{r_0} \right)^{-\delta},
    \end{equation}
    where $\gamma$ and $\delta$ are the logarithmic slopes of the three-dimensional mass and luminosity distributions, respectively.
    \item The stellar orbital anisotropy, $\beta = 1 - \sigma_\theta^2/\sigma_r^2$, is constant with radius. Here, $\sigma_r$ and $\sigma_\theta$ are the radial and tangential velocity dispersion components, respectively.
\end{enumerate}

\noindent The projected dynamical mass within a cylinder of radius $R_{\mathrm{E}} = D_l \theta_{\mathrm{E}}$ is obtained by integrating the surface mass density for the power-law model in Equation~\eqref{eq:rho}:
\begin{equation} \label{eq:Mdyn_proj}
    M_{\mathrm{dyn}}^{\mathrm{E}} = \frac{2\pi^{3/2}}{3-\gamma} \,
    \frac{\rho_0}{r_0^{-\gamma}} \,
    R_{\mathrm{E}}^{3-\gamma} \,
    \frac{\Gamma\!\left(\frac{\gamma-1}{2}\right)}{\Gamma\!\left(\frac{\gamma}{2}\right)},
\end{equation}
where $\Gamma$ denotes the Gamma function.

The stellar dynamics are governed by the spherical Jeans equation. For the power-law profiles and constant anisotropy $\beta$, the solution relates the radial velocity dispersion to the enclosed mass and luminosity. The observable quantity is the luminosity-weighted, aperture-averaged line-of-sight velocity dispersion. For comparison with spectroscopic data, we compute its value within a standardized aperture of radius $\theta_{\mathrm{eff}}/2$ (half the effective radius). Combining the solution of the Jeans equation with the expression for $M_{\mathrm{dyn}}^{\mathrm{E}}$ yields the key theoretical prediction (see \citealt{Chen2019} for a complete derivation):
\begin{equation} \label{eq:sigma_th}
    \sigma_{\parallel}^{\mathrm{th}}(\leq \theta_{\mathrm{eff}}/2)
    = \sqrt{
        \frac{c^2}{2\sqrt{\pi}} \,
        \frac{D_s}{D_{ls}} \,
        \theta_{\mathrm{E}} \,
        f(\gamma, \delta, \beta) \,
        \left( \frac{\theta_{\mathrm{eff}}}{2\theta_{\mathrm{E}}} \right)^{2-\gamma}
    },
\end{equation}
where the dimensionless factor $f(\gamma, \delta, \beta)$ encapsulates the integrated dynamical effects:
\begin{equation} \label{eq:f_factor}
\begin{aligned}
    f(\gamma, \delta, \beta)
    &= \frac{3-\delta}{(\xi - 2\beta)(3-\xi)} \,
      \left[ \frac{\Gamma\!\left(\frac{\xi-1}{2}\right)}{\Gamma\!\left(\frac{\xi}{2}\right)}
           - \beta \frac{\Gamma\!\left(\frac{\xi+1}{2}\right)}{\Gamma\!\left(\frac{\xi+2}{2}\right)} \right] \\
    &\quad \times \frac{\Gamma\!\left(\frac{\gamma}{2}\right) \Gamma\!\left(\frac{\delta}{2}\right)}
                      {\Gamma\!\left(\frac{\gamma-1}{2}\right) \Gamma\!\left(\frac{\delta-1}{2}\right)},
    \quad \text{with} \quad \xi = \gamma + \delta - 2.
\end{aligned}
\end{equation}
Equation~\eqref{eq:sigma_th} provides the critical link between lensing observables ($\theta_{\mathrm{E}}$, $D_s/D_{ls}$), galaxy structural parameters ($\theta_{\mathrm{eff}}$, $\delta$), the mass-profile slope $\gamma$, and the stellar anisotropy $\beta$.

\subsection{Aperture correction and observables}

The observed velocity dispersion $\sigma_{\mathrm{ap}}$ is measured within an instrumental aperture $\theta_{\mathrm{ap}}$. To compare with the model prediction $\sigma_{\parallel}^{\mathrm{th}}(\leq \theta_{\mathrm{eff}}/2)$, we apply an aperture correction to standardize the measurement:
\begin{equation}
\label{eq:aperture_correction}
\sigma_{\mathrm{e2}}^{\mathrm{obs}} \equiv \sigma_{\parallel}^{\mathrm{obs}}(\leq \theta_{\mathrm{eff}}/2) =
\sigma_{\mathrm{ap}} \left[ \frac{\theta_{\mathrm{eff}}}{2\theta_{\mathrm{ap}}} \right]^{\eta},
\end{equation}
where the exponent $\eta$ accounts for the radial gradient of the velocity dispersion. We adopt the prior $\eta = -0.066 \pm 0.035$ from \citet{Cappellari2006} and \citet{Chen2019}.

\subsection{Statistical framework and parameter inference}

The joint analysis is performed by comparing, for each lens system, the theoretically predicted velocity dispersion (Equation~\ref{eq:sigma_th}) with the aperture-corrected observed value (Equation~\ref{eq:aperture_correction}). This comparison is encoded in a Gaussian likelihood function:
\begin{equation}
\label{eq:likelihood}
    \mathcal{L} \propto \exp\left(-\frac{\chi^2}{2}\right), \quad \text{with} \quad \chi^2 = \sum_{i=1}^{N} \left( \frac{\sigma_{\parallel, i}^{\mathrm{th}} - \sigma_{\mathrm{e2}, i}^{\mathrm{obs}}}{\Delta\sigma_{\mathrm{tot}, i}} \right)^2.
\end{equation}
The total uncertainty $\Delta\sigma_{\mathrm{tot}}$ for each lens incorporates the measurement error of $\sigma_{\mathrm{ap}}$, the uncertainty propagated from the aperture correction ($\eta$), and a systematic term accounting for potential effects like line-of-sight mass contamination.

We adopt a Bayesian approach to infer the posterior probability distributions of the model parameters. The primary parameters of interest are those governing the mass-density slope $\gamma$. In this work, we parameterize $\gamma$ to investigate its evolution with redshift and dependence on galaxy surface density. The stellar luminosity density slope $\delta$ and the velocity anisotropy $\beta$ are treated as global nuisance parameters. We marginalize over them using informed Gaussian priors: $\delta = 2.173 \pm 0.085$ and $\beta = 0.18 \pm 0.13$, consistent with previous studies of massive ETGs \citep{Chen2019}. Parameter estimation is performed via Monte Carlo nested sampling using the \texttt{PyMultiNest} package \citep{Buchner2014}.

The joint lensing and dynamics analysis constrains the mass-density slope $\gamma$ directly by equating the lensing and dynamical masses within the Einstein radius. We adopt a power-law model for the total mass density and parameterise its logarithmic slope to study its evolution with redshift and its dependence on galaxy properties, following \citet{Chen2019}:
\begin{equation}
\label{eq:gamma_param}
\gamma = \gamma_0 + \gamma_z \cdot z_l + \gamma_s \cdot \log \widetilde{\Sigma}.
\end{equation}
Here, $\log$ denotes the base-10 logarithm, $\gamma_0$ is a reference slope, $\gamma_z$ quantifies the redshift evolution (capturing time-dependent structural changes), and $\gamma_s$ measures the dependence on the surface mass density $\widetilde{\Sigma}$, defined as
\begin{equation}
\widetilde{\Sigma} = \frac{(\sigma_{\mathrm{e2}} / 100\ \text{km s}^{-1})^2}{R_{\mathrm{eff}} / 10\ h^{-1}\ \text{kpc}},
\end{equation}
where $\sigma_{\mathrm{e2}}$ is the aperture-corrected velocity dispersion taken from Equation~(\ref{eq:aperture_correction}), $R_{\mathrm{eff}}$ is the effective radius in physical units, and the Hubble constant follows the usual convention $H_0 = 100h\ \text{km s}^{-1}\ \text{Mpc}^{-1}$.

The predicted aperture velocity dispersion (Equation~\ref{eq:sigma_th}) depends on cosmological distances. We work within the flat $\Lambda$CDM framework, with angular diameter distances given by
\begin{align}
D_s(z_s; \Omega_m, H_0) &= \frac{c}{H_0 (1+z_s)} \int_0^{z_s} \frac{dz'}{E(z'; \Omega_m)}, \label{Dl} \\
D_{ls}(z_l, z_s; \Omega_m, H_0) &= \frac{c}{H_0 (1+z_s)} \int_{z_l}^{z_s} \frac{dz'}{E(z'; \Omega_m)}, \label{Dls}
\end{align}
where the expansion function is
\begin{equation} \label{Ezp}
    E(z; \Omega_m) = \sqrt{\Omega_{m}(1+z)^3 + (1 - \Omega_{m})}.
\end{equation}
A key advantage of this method is the cancellation of the Hubble constant $H_0$ in the ratio $D_s/D_{ls}$, making our results robust to its current observational uncertainties.

We perform a Bayesian analysis to infer the posterior distributions of the model parameters. The full parameter vector is $\boldsymbol{\Theta} = \{\gamma_0, \gamma_z, \gamma_s, \Omega_m\}$. The stellar luminosity density slope $\delta$ and the velocity anisotropy $\beta$ are treated as global nuisance parameters, marginalised over using informed Gaussian priors: $\delta = 2.173 \pm 0.085$ and $\beta = 0.18 \pm 0.13$ \citep{Chen2019}. For the matter density parameter, we adopt a tight Gaussian prior from the latest DESI BAO measurements: $\Omega_m = 0.2975 \pm 0.0086$ \citep{DESI2025_DR2_BAO}. Broad, uninformative flat priors are assigned to the parameters of interest: $\gamma_0 \in [0.1, 3.0]$, $\gamma_z \in [-2.0, 2.0]$, and $\gamma_s \in [-2.0, 2.0]$.

\section{Results and Discussion}
\label{sect:analysis}

\begin{figure*}
    \centering
    \includegraphics[width=0.7\linewidth]{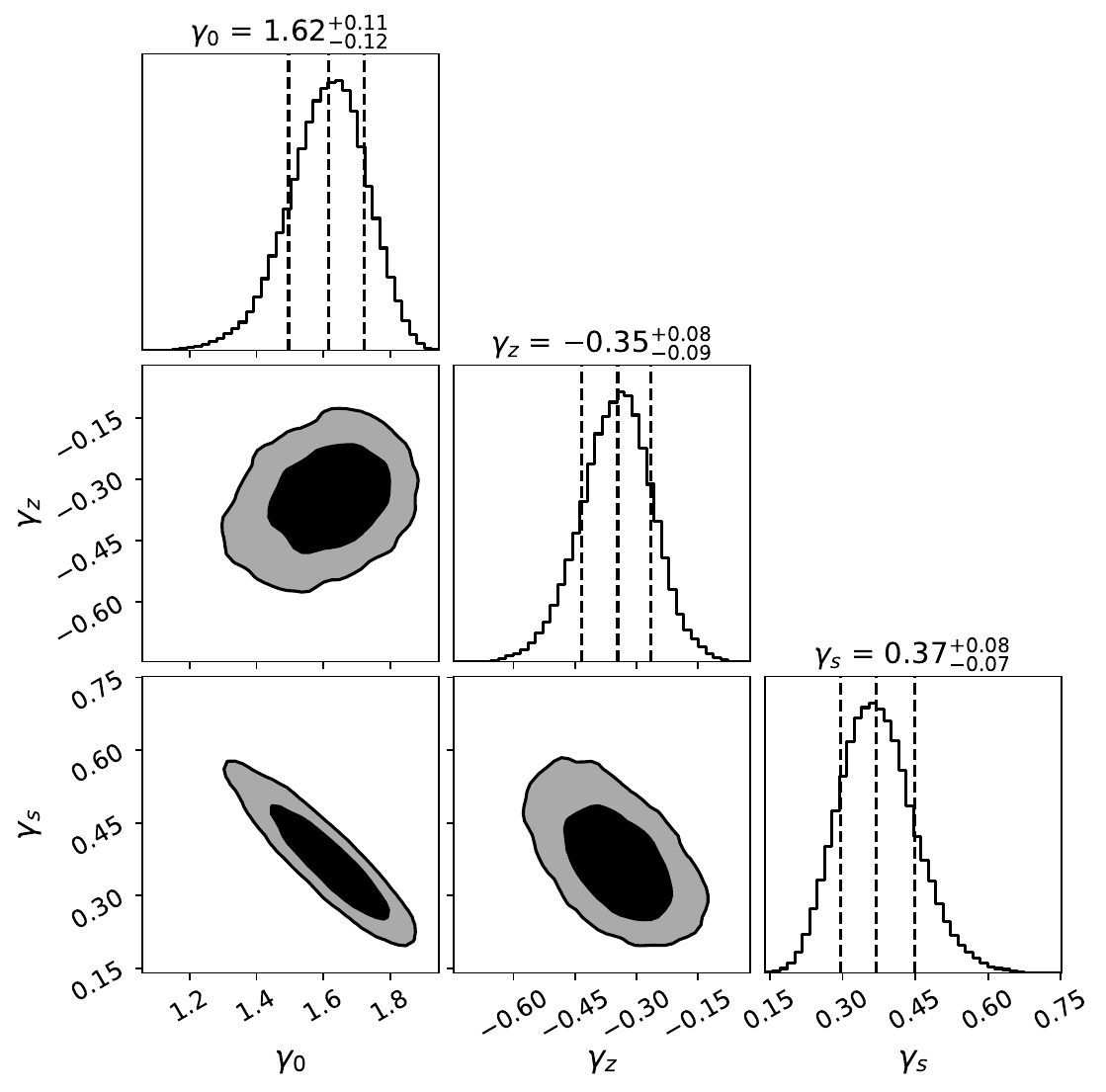}
    \caption{Posterior probability distributions for the mass density slope parameters $\gamma_0$, $\gamma_z$, and $\gamma_s$. Diagonal panels show the one-dimensional marginalised posterior distributions, with dashed lines indicating the 68\% credible intervals and the mean value. Off-diagonal panels show the two-dimensional joint posterior distributions, with contours marking the 68\% and 95\% credible regions. The constraints, derived from the joint analysis of 24 lensed quasar systems, reveal the dependence of the slope on redshift ($\gamma_z$) and surface mass density ($\gamma_s$).}
    \label{fig:posteriors}
\end{figure*}

Figure~\ref{fig:posteriors} shows the marginalised posterior distributions for the key parameters $\gamma_0$, $\gamma_z$, and $\gamma_s$. From the analysis of our 24 lensed quasar systems, we obtain the following constraints (68\% credible intervals):
\begin{align}
    \gamma_0 &= 1.62^{+0.11}_{-0.12}, \\
    \gamma_z &= -0.35^{+0.08}_{-0.09}, \\
    \gamma_s &= 0.37^{+0.08}_{-0.07}.
\end{align}
The intercept $\gamma_0 = 1.62$ is significantly shallower than the isothermal value of 2. The negative $\gamma_z$ indicates a clear trend: the mass density profile becomes shallower (lower $\gamma$) at higher redshifts. Conversely, the positive $\gamma_s$ reveals a strong correlation: at fixed redshift, galaxies with higher surface mass density $\widetilde{\Sigma}$ possess steeper mass profiles.

Our results are in good qualitative agreement with major studies of GGSL \citep{Auger2010, Bolton2012, Chen2019}. The inferred trends—negative evolution with redshift and positive dependence on surface density—support the picture of inside-out growth and mass assembly of massive early-type galaxies within the $\Lambda$CDM paradigm.

Quantitatively, the evolutionary trend measured in this work ($\gamma_z = -0.35^{+0.08}_{-0.09}$) is stronger than that reported by \citet{Chen2019} for GGSL systems ($\gamma_z = -0.218^{+0.089}_{-0.087}$). Although the final sample employed in our joint analysis is approximately one-fifth the size of that used in Table P3 of \citet{Chen2019}, the associated uncertainties do not scale according to the expected $\sqrt{5}$ relation. 
This difference is likely driven by a combination of methodological and sample‑selection effects. In particular, following \citet{Chen2019}, our analysis adopts informative Gaussian priors on the nuisance parameters ($\delta = 2.173 \pm 0.085$ and $\beta = 0.18 \pm 0.13$), which naturally contributes to tighter posterior constraints. 
To quantify this effect, we re‑analysed the 130 systems from \citet{Chen2019} using the same prior setup adopted in this work. We found that the uncertainty on $\gamma_z$ decreases from $0.088$, as reported in \citet{Chen2019}, to approximately $0.07$. This confirms that the informative priors do contribute to tightening the inferred constraint. Nevertheless, the uncertainty obtained in the present work, $\Delta \gamma_z \simeq 0.09$, remains considerably smaller than would be expected from a simple sample-size scaling from 130 to 24 systems, indicating that the difference in the resulting uncertainties is not driven solely by the adoption of informative priors.
Rather, the marked difference in the lens redshift coverage between the two samples appears to be the primary contributor. 
The lens redshift range in \citet{Chen2019} spans $0.062 \leq z_l \leq 0.722$, whereas our GQSL sample spans $0.195 \leq z_l \leq 0.867$. In addition to this broader coverage of the high-redshift regime, our lens population contains a substantially higher fraction of systems at $z_l > 0.4$ ($66.7\%$) compared to the \citet{Chen2019} compilation ($27.7\%$).
This is a natural consequence of the typically higher source redshifts of background quasars, which allow foreground lens galaxies to span a broader and generally higher-redshift interval,
and thus highlight the unique potential of GQSL for probing the high-redshift regime. The inferred dependence on surface mass density ($\gamma_s \approx 0.37$) is consistent with earlier constraints, albeit slightly shallower, thereby reinforcing the view that massive early-type galaxies share fundamental structural properties irrespective of the nature of the background source used to probe them.

Another point worth discussing is that our analysis adopts a spherical assumption, whereas the Einstein radii used in this work were originally derived from non‑spherical lens models, mainly SIE and power‑law models. For our analysis sample, the mean axis ratio is approximately $q \approx 0.75$, indicating that these lens systems are not highly flattened. In this context, the key issue is not simply the geometric difference between an ellipse and a circle, but rather how much the projected mass enclosed within a circular aperture of radius $R_E$ deviates from the spherical expression $M_{\mathrm{grl}}^{E} = \Sigma_{\mathrm{cr}} \pi R_E^2$, given that the adopted Einstein radius originates from an elliptical lens model. For many commonly used elliptical models, the Einstein radius is itself defined through an equivalent enclosed mass, so the corresponding projected mass remains consistent with the spherical expression. For the remaining cases, based on our sample's mean axis ratio of $q \approx 0.75$, we estimate that the difference between the projected mass enclosed within a circular aperture of radius $R_E$ and that in the spherical case is only about $0.5\%$. The resulting impact on the inferred density‑slope parameter is expected to be even smaller and remains well below the current statistical uncertainties. This indicates that the final uncertainties on the mass‑profile slope and other key parameters are sufficiently large to absorb this effect at the level of the present analysis. Nevertheless, this estimate only accounts for the geometric mismatch on the lensing‑mass side; the spherical assumption in the dynamical modelling itself may introduce additional residual systematic uncertainty, which deserves a more complete treatment in future work.

Previous joint strong‑lensing and stellar‑dynamical studies have often reported a mild tendency for the inferred total density slope to become steeper towards lower redshift \citep{Ruff2011, Bolton2012, Sonnenfeld2013b}. One possible interpretation is that dissipative baryonic processes and structural evolution increase the central mass concentration, consistent with the empirical finding that more compact galaxies, or those with higher stellar mass surface density, tend to have steeper total density slopes. At the same time, this trend should be interpreted with caution, since several studies have argued that part of the apparent redshift dependence may arise from selection effects and modelling assumptions, in particular the adoption of a power‑law mass profile and simplified treatments of stellar orbital anisotropy \citep{Shankar2017, Xu2017}. This issue becomes especially important when comparing with cosmological simulations. 
Several cosmological hydrodynamical simulations predict that the intrinsic total density slopes of massive early-type galaxies are steeper at higher redshift and evolve only weakly below $z \sim 1$, although the sign and amplitude of the trend depend on the adopted feedback model, radial range, and sample selection \citep{Johansson2012, Remus2013, Remus2017, Xu2017, Wang2020, Derkenne2021}. This behaviour is commonly associated with the decreasing importance of dissipative processes towards low redshift, late-time dry mergers, size growth, increasing central dark matter fractions, and AGN feedback, all of which can make the inner total mass distribution closer to isothermal or slightly shallower. However, when simulated galaxies are analysed using observationally motivated lensing–dynamical methods, part of the observationally inferred steepening towards lower redshift can be recovered \citep{Xu2017, Remus2017}.
Therefore, while our result is broadly consistent with the observational strong‑lensing and dynamics literature, the comparison with simulations suggests that the observed trend may reflect a combination of genuine physical evolution and non‑negligible observational or modelling systematics.

This study successfully applies the joint lensing and dynamics method to a substantial sample of lensed quasars for the first time. It demonstrates that robust constraints on galaxy mass profiles can be obtained despite the challenge of contamination from bright background sources, thereby providing a complementary probe of galaxy structure and evolution.

\section{Conclusions}
\label{sect:conclusion}

We have carried out a systematic investigation of the internal mass structure of ETGs through the construction and analysis of a well-defined sample of galaxy-scale strong gravitational lenses with background quasars. For the lens probability statistics \citep{1984ApJ...284....1T,Kochanek1992,2003MNRAS.343..639O,Cao2012,Li2023}  that rely solely on lens and source redshifts and Einstein radii, the full parent sample of 106 systems is already suitable and provides a robust foundation. From this parent sample, we carefully selected a subset of 24 systems with the ancillary data necessary for joint lensing and dynamical modelling—specifically, stellar velocity dispersions and effective radii. The assembly and rigorous characterisation of this subset, including the acquisition of 11 new stellar velocity dispersions via an iterative spectroscopic fitting technique designed to suppress quasar contamination, represents a cornerstone of the present work. By jointly analysing precise strong-lensing geometry and stellar kinematics within a unified framework, we have placed robust constraints on the slope of the total mass-density profile and its evolution, thereby demonstrating the power of GQSL as a complementary probe of galaxy structure.

Our principal results are summarised as follows:
\begin{enumerate}
    \item Modelling the total three-dimensional mass-density profile of the lens galaxies as a power law, $\rho \propto r^{-\gamma}$, we parameterised its logarithmic slope as $\gamma = \gamma_0 + \gamma_z \cdot z_l + \gamma_s \cdot \log \tilde{\Sigma}$, where $z_l$ is the lens redshift and $\tilde{\Sigma}$ is the surface mass density. Assuming a flat $\Lambda$CDM cosmology and employing the latest DESI BAO measurements as a prior, we constrained the parameters to $\gamma_0 = 1.62^{+0.11}_{-0.12}$, $\gamma_z = -0.35^{+0.08}_{-0.09}$, and $\gamma_s = 0.37^{+0.08}_{-0.07}$ (68\% confidence intervals).
    \item The negative value of $\gamma_z$ indicates a significant evolutionary trend: the mass-density profiles of ETGs become systematically shallower (i.e., have a lower $\gamma$) at higher redshifts. Conversely, the positive value of $\gamma_s$ reveals a strong positive correlation with surface mass density: at a fixed redshift, galaxies with denser stellar cores possess steeper mass distributions.
    \item These findings are qualitatively consistent with previous studies based on GGSL systems, reinforcing the established picture of mass-structure evolution in massive ETGs. 
    Quantitatively, the stronger redshift evolution we measure is likely attributable to the distinct composition of our GQSL sample. In particular, our sample spans $0.195 \leq z_l \leq 0.867$, compared to $0.062 \leq z_l \leq 0.722$ in \citet{Chen2019}. In addition, our sample contains a substantially higher fraction of high-redshift lenses, with $66.7\%$ of systems at $z_l > 0.4$, compared to $27.7\%$ in \citet{Chen2019}. This demonstrates the value of our sample in providing an independent and complementary constraint on structural changes over cosmic time.
\end{enumerate}

The principal novel contributions of this work are threefold: (i) the assembly and rigorous characterisation of a substantial sample of galaxy-scale strong lenses with background quasars, specifically compiled for joint lensing and dynamical analysis; (ii) the acquisition of new stellar velocity dispersion measurements for 11 lens galaxies and effective radii for 4 systems from archival data, achieved through a robust iterative spectroscopic fitting technique that effectively mitigates quasar contamination—this effort expanded the subset of systems amenable to our analysis from 13 to 24; and (iii) the first successful application of the joint lensing–dynamics framework to a sizeable population of lensed quasars, thereby establishing this class of systems as an independent and complementary probe of galaxy structure.

This study has several limitations. Whilst substantial, the current sample size limits the statistical precision of our constraints. The adopted power-law mass model, although effective for the scales probed, may not capture the full structural complexity of galaxies. Furthermore, our inference relies on priors for the nuisance parameters $\delta$ and $\beta$, which are derived from external studies.
Looking ahead, the inclusion of extended arc data would provide substantially richer lensing constraints, enabling a more refined and realistic reconstruction of the mass distribution and helping to reduce modelling degeneracies, particularly in constraining the mass profile. Another limitation of the present analysis is the use of single‑fiber velocity dispersion measurements, which – although valuable – are restricted by their limited spatial resolution. In the future, spatially resolved kinematic data will yield much stronger constraints on the internal mass distribution by exploiting the full spatial information available in spectroscopy. Such data will also help to mitigate the impact of quasar contamination on the measurements.

Future prospects in this field are highly promising. Upcoming deep, wide-field imaging and spectroscopic surveys, such as the Chinese Space Station Survey Telescope (CSST) \citep{CSST:2025ssq,2025arXiv251108030W,2024MNRAS.533.1960C}, the Vera C. Rubin Observatory \citep{2019ApJ...873..111I}, the \textit{Euclid} mission \citep{2022A&A...662A.112E}, and DESI \citep{2025arXiv250918089H}, are expected to yield a dramatic increase in the number of discovered and well-characterised strong lens systems. Larger, more homogeneous samples of lensed quasars will enable more precise constraints on mass-profile evolution, permit the exploration of additional dependencies (e.g., on environment or morphology), and strengthen the foundation for using strong lensing as a simultaneous probe of galaxy astrophysics and cosmology. The spectroscopic methodology developed here to handle bright quasar contamination will be directly applicable and valuable for analysing these future large samples.

\section*{Acknowledgements}
This work was supported by the National Key Research and Development Programme of China (Grant Nos. 2022YFA1602903 and 2023YFB3002501), the National Natural Science Foundation of China (Grant Nos. 12403104, 12473002, 12588202, and 12165021), and the China Manned Space Programme (Grant No. CMS-CSST-2025-A03). J. W. gratefully acknowledges the support from the Young Talent Programme under the Xingdian Talent Support Plan.

\section*{Data Availability}
The Gravitationally
Lensed Quasar Database is available at \url{https://research.ast.cam.ac.uk/lensedquasars/index.html}. Model posterior chains are available from the corresponding author on request.

\bibliographystyle{mnras}
\bibliography{example} 

\onecolumn
\appendix
\section{Parent Sample}
\begin{longtable}{>{\raggedright\arraybackslash}p{2.5cm} >{\centering\arraybackslash}p{1.5cm} >{\centering\arraybackslash}p{1.5cm} >{\centering\arraybackslash}p{0.8cm} >{\centering\arraybackslash}p{0.8cm} >{\centering\arraybackslash}p{0.8cm}>{\centering\arraybackslash}p{0.15cm} >{\centering\arraybackslash}p{6.9cm}}
    \caption{The parent sample of 106 confirmed galaxy-scale strong gravitational lenses with background quasars.} \label{all samples}\\
   
    \hline
    \noalign{\smallskip}
    Name & RA & Dec & $z_l$ & $z_s$ & $\theta_{\text{max}}$ & $N_{\text{img}}$ & Reference\\
         & (deg) & (deg) &   &  & (arcsec) &  & \\
    \hline
    \noalign{\smallskip}
    \endfirsthead
    \hline
    \multicolumn{7}{|c|}{\textit{Table \thetable\ (Continued)}} \\
    \hline
    Name & RA & Dec & $z_l$ & $z_s$ & $\theta_{\text{max}}$ & $N_{\text{img}}$ & Reference\\
         & (deg) & (deg) &   &  & (arcsec) &  & \\
    \hline
    \noalign{\smallskip}
    \endhead
    \hline
    Name & RA & Dec & $z_l$ & $z_s$ & $\theta_{\text{max}}$ & $N_{\text{img}}$ & Reference\\
         & (deg) & (deg) &   &  & (arcsec) &  & \\
    \hline
    \noalign{\smallskip}
    \endhead
    \endfoot
    \hline
    \noalign{\smallskip}
    \endlastfoot
J0030-3358 & 7.6740 & -33.9767 & 0.715 & 1.58 & 2.03 & 2 & \citet{Lemon2023} \\
HE0047-1756 & 12.6158 & -17.6693 & 0.408 & 1.67 & 1.44 & 2 & \citet{Wisotzki2004, Ofek2006} \\
SDSSJ0114+0722 & 18.65991 & 7.37458 & 0.408 & 1.828 & 1.70 & 2 & \citet{More2016} \\
A0140-1152 &  25.012499 & -11.871944 & 0.277 & 1.805 & 1.46 & 2 & \citet{Agnello2018b} \\
Q0142-100 & 26.3194 & -9.75475 & 0.491 & 2.719 & 2.22 & 2 & \citet{Surdej1988} \\
PSJ0147+4630 & 26.79231 & 46.5118 & 0.678 & 2.341 & 3.8 & 4 & \citet{Berghea2017, Lee2018} ;\\
 &  &  &  &  &  &  & \citet{Goicoechea2019}\\
QJ0158-4325 & 29.6724 & -43.4178 & 0.317 & 1.294 & 1.222 & 2 & \citet{Morgan1999, Faure2009}\\
J0203+1612 & 30.9977 & 16.20213 & 0.488 & 2.18 & 2.73 & 2 & \citet{Lemon2019} \\
SDSSJ0246-0825 & 41.64204 & -8.42669 & 0.724 & 1.68 & 1.04 & 2 & \citet{Inada2005} \\
J0247-0800 & 41.9561 & -8.0150 & 0.198 & 3.28 & 1.68 & 2 & \citet{Lemon2023} \\
SDSSJ0256+0153 & 44.16983 & 1.89147 & 0.603 & 2.600 & 1.93 & 2 & \citet{More2016} \\
J0310-5545 & 47.7029 & -55.7534 & 0.298 & 2.31 & 3.57 & 2 & \citet{Lemon2023} \\
J0347-2154 & 56.7690 & -21.9095 & 0.187 & 0.81 & 1.87 & 2 & \citet{Lemon2023} \\
MG0414+0534 & 63.6572 & 5.5786 & 0.96 & 2.64 & 2.4 & 4 & \citet{Hewitt1992, Lawrence1995} ;\\
 &  &  &  &  &  &  & \citet{Tonry1999}\\
J0416+7428 & 64.1972 & 74.4827 & 0.098 & 0.900 & 2.64 & 2 & \citet{Lemon2023} \\
HE0435-1223 & 69.5619 & -12.28745 & 0.454 & 1.689 & 2.6 & 4 & \citet{Wisotzki2002, Eigenbrod2006} \\
HE0512-3329 & 78.54541 & -33.43958 & 0.9313 & 1.565 & 0.644 & 2 & \citet{Gregg2000} \\
J0635+6452 & 98.9864 & 64.8715 & 0.427 & 1.845 & 3.07 & 2 & \citet{Lemon2023} \\
J0643+2725 & 100.9259 & 27.4276 & 0.185 & 1.562 & 2.44 & 2 & \citet{Lemon2023} \\
J0659+1629 & 104.76823 & 16.485772 & 0.766 & 3.09 & 6.8 & 4 & \citet{Delchambre2019, Stern2021} \\
SDSSJ0743+2457 & 115.9692 & 24.96211 & 0.381 & 2.165 & 1.034 & 2 & \citet{Jackson2012, Inada2014} \\
SDSSJ0746+4403 & 116.721 & 44.06425 & 0.513 & 1.998 & 1.08 & 2 & \citet{Inada2007, Kayo2010} \\
SDSSJ0806+2006 & 121.59867 & 20.10874 & 0.573 & 1.540 & 1.40 & 2 & \citet{Inada2006} \\
J0818-2613 & 124.6179 & -26.2236 & 0.866 & 2.164 & 6.2 & 4 & \citet{Stern2021, Mozumdar2023} \\
SDSSJ0819+5356 & 124.99909 & 53.94018 & 0.294 & 2.237 & 4.04 & 2 & \citet{Inada2009} \\
ULASJ0820+0812 & 125.06706 & 8.20466 & 0.803 & 2.024 & 2.3 & 2 & \citet{Jackson2009} \\
HS0818+1227 & 125.4121 & 12.29191 & 0.39 & 3.115 & 2.1 & 2 & \citet{Hagen2000} \\
SDSSJ0821+4542 & 125.4944 & 45.7123 & 0.349 & 2.066 & 1.35 & 2 & \citet{More2016} \\
SDSSJ0832+0404 & 128.071 & 4.06792 & 0.659 & 1.115 & 1.98 & 2 & \citet{Oguri2008} \\
J0840+3550 & 130.13842 & 35.83334 & 0.26 & 1.77 & 2.46 & 2 & \citet{Lemon2018} \\
B0850+054 & 133.22323 & 5.25435 & 0.59 & 1.14 & 0.7 & 2 & \citet{Biggs2003} \\
SDSSJ0903+5028 & 135.89609 & 50.47213 & 0.388 & 3.6 & 2.8 & 2 & \citet{Johnston2003} \\
J0911-0948 & 137.7845 & -9.8054 & 0.251 & 1.47 & 2.52 & 2 & \citet{Lemon2023} \\
SBS0909+532 & 138.25425 & 52.99133 & 0.830 & 1.388 & 1.107 & 2 & \citet{Kochanek1997, Lubin2000}\\
J0918-0220 & 139.6806 & -2.3354 & 0.460 & 0.803 & 2.26 & 2 & \citet{Jaelani2021, Lemon2023} \\
J0921+3020 & 140.2685 & 30.3421 & 0.428 & 3.335 & 2.93 & 2 & \citet{Lemon2023} \\
SDSSJ0921+2854 & 140.3144 & 28.9123 & 0.445 & 1.410 & 1.91 & 2 & \citet{More2016} \\
J0924+4235 & 141.1243 & 42.5947 & 0.415 & 3.17 & 4.64 & 2 & \citet{Lemon2023} \\
SDSSJ0924+0219 & 141.23246 & 2.32358 & 0.393 & 1.524 & 1.78 & 4 & \citet{Inada2003, Ofek2006}\\
J0936-1211 & 144.2494 & -12.1836 & 0.260 & 2.00 & 2.11 & 2 & \citet{Lemon2023} \\
J0941+0518 & 145.34378 & 5.30664 & 0.34 & 1.54 & 5.40 & 2 & \citet{Lemon2018} \\
SDSSJ0946+1835 & 146.52017 & 18.59453 & 0.388 & 4.8 & 3.06 & 2 & \citet{McGreer2010} \\
J0949+4208 & 147.47830 & 42.13381 & 0.51 & 1.27 & 2.57 & 2 & \citet{Lemon2018} \\
BRI0952-0115 & 148.75038 & -1.50187 & 0.632 & 4.50 & 0.9 & 2 & \citet{McMahon1992, Eigenbrod2006} \\
J0957-2242 & 149.469625 & -22.700822 & 0.492 & 2.061 & 0.670 & 4 & \citet{Scialpi2025} \\
COSMOS5921+0638 & 149.84071 & 2.11063 & 0.551 & 3.14 & 1.4 & 4 & \citet{Faure2008, Anguita2009} \\
Q0957+561 & 150.33665 & 55.89791 & 0.3562 & 1.413 & 6.16 & 2 & \citet{walsh1979, Chartas1998};\\
 &  &  &  &  &  &  & \citet{Krips2005}\\
SDSSJ1001+5027 & 150.36876 & 50.46595 & 0.415 & 1.838 & 2.86 & 2 & \citet{Oguri2005, Misawa2018} \\
CXCOJ1002+0203 & 150.5063 & 2.0581 & 0.439 & 2.016 & 0.85 & 2 & \citet{Jaelani2021} \\
J1003+0651 & 150.7886 & 6.8501 & 0.225 & 2.565 & 2.62 & 2 & \citet{Lemon2023} \\
FIRSTJ1004+1229 & 151.10379 & 12.48966 & 0.95 & 2.65 & 1.5 & 2 & \citet{Lacy2002} \\
Q1009-0252 & 153.06622 & -3.1176 & 0.869 & 2.739 & 1.53 & 2 & \citet{Hewett1994} \\
SDSSJ1021+4913 & 155.29588 & 49.22508 & 0.451 & 1.72 & 1.14 & 2 & \citet{Pindor2006, Inada2012}  \\
FSC10214+4724 & 156.14392 & 47.15272 & 0.896 & 2.286 & 2.53 & 2 & \citet{RowanRobinson1991, Serjeant1995} \\
B1030+074 & 158.39177 & 7.19 & 0.599 & 1.535 & 1.56 & 2 & \citet{Xanthopoulos1998} \\
W2MJ1042+1641 & 160.59213 & 16.68758 & 0.599 & 2.517 & 1.8 & 4 &  \citet{Glikman2023}\\
SDSSJ1055+4628 & 163.93938 & 46.4777 & 0.388 & 1.249 & 1.15 & 2 & \citet{Kayo2010, Rusu2016} \\
HE1104-1805 & 166.63938 & -18.35672 & 0.729 & 2.319 & 3.19 & 2 & \citet{Wisotzki1993} \\
PG1115+080 & 169.57035 & 7.76627 & 0.311 & 1.722 & 2.26 & 4 & \citet{Weymann1980, Kundic1997} \\
RXJ1131-1231 & 172.9644 & -12.5329 & 0.295 & 0.658 & 3.1 & 4 & \citet{Sluse2003} \\
SDSSJ1138+0314 & 174.51554 & 3.24939 & 0.445 & 2.438 & 1.46 & 4 & \citet{Eigenbrod2006} \\
SDSSJ1155+6346 & 178.82195 & 63.77265 & 0.1756 & 2.89 & 1.832 & 2 & \citet{Pindor2004} \\
B1152+199 & 178.82639 & 19.66146 & 0.4386 & 1.0189 & 1.56 & 2 & \citet{Myers1999} \\
SDSSJ1206+4332 & 181.62361 & 43.53856 & 0.748 & 1.789 & 2.90 & 2 & \citet{Oguri2005, Birrer2019} \\
Q1208+1011 & 182.73764 & 9.90749 & 1.1349 & 3.803 & 0.45 & 2 & \citet{Magain1992, Siemiginowska1998} \\
HSCJ1215-0058 & 183.88950 & -0.97856 & 0.45912 & 2.8817 & 4.92 & 2 & \citet{Shu2025, He2025} \\
HSCJ1220+0112 & 185.079002 & 12.15165 & 0.48755 & 1.7081 & 1.87 & 2 & \citet{Shu2025, He2025} \\
SDSSJ1226-0006 & 186.5334 & -0.10061 & 0.517 & 1.212 & 1.24 & 2 & \citet{Inada2008, Eigenbrod2006} \\
SDSSJ1251+2935 & 192.78154 & 29.59458 & 0.410 & 0.802 & 1.79 & 4 & \citet{Kayo2007} \\
SDSSJ1254+1857 & 193.6682 & 18.9533 & 0.555 & 1.717 & 2.32 & 2 & \citet{More2016} \\
J1303+1816 & 195.7765 & 18.2781 & 0.46 & 2.95 & 2.26 & 2 & \citet{Lemon2023} \\
SDSSJ1313+5151 & 198.41675 & 51.85801 & 0.194 & 1.875 & 1.24 & 2 & \citet{Ofek2007} \\
J1326+3020 & 201.7410 & 30.3400 & 0.339 & 1.852 & 2.11 & 2 & \citet{Lemon2023} \\
SDSSJ1330+1810 & 202.57772 & 18.17581 & 0.373 & 1.393 & 1.76 & 4 & \citet{Oguri2008} \\
SDSSJ1332+0347 & 203.09425 & 3.79442 & 0.191 & 1.445 & 1.14 & 2 & \citet{Morokuma2007} \\
SDSSJ1334+3315 & 203.50579 & 33.25953 & 0.557 & 2.426 & 0.833 & 2 & \citet{Rusu2011} \\
SDSSJ1335+0118 & 203.89496 & 1.30153 & 0.440 & 1.57 & 1.56 & 2 & \citet{Oguri2004} \\
SDSSJ1339+1310 & 204.77999 & 13.1775 & 0.607 & 2.243 & 1.69 & 2 & \citet{Inada2009} \\
Q1355-2257 & 208.9308 & -22.9565 & 0.701 & 1.37 & 1.22 & 2 & \citet{Morgan2003} \\
SDSSJ1406+6126 & 211.60303 & 61.44446 & 0.27 & 2.13 & 1.98 & 2 & \citet{Inada2007} \\
HST1411+5211 & 212.8317 & 52.19158 & 0.465 & 2.811 & 1.8 & 4 & \citet{Fischer1998, Lubin2000} \\
B1422+231 & 216.15871 & 22.9335 & 0.339 & 3.62 & 1.28 & 4 & \citet{Patnaik1992, Hammer1995} \\
SDSSJ1433+6007 & 218.345 & 60.121 & 0.407 & 2.737 & 3.6 & 4 & \citet{Agnello2018} \\
SDSSJ1452+4224 & 223.0479 & 42.4082 & 0.382 & 4.819 & 1.59 & 2 & \citet{More2016} \\
SDSSJ1515+1511 & 228.9107 & 15.19327 & 0.742 & 2.054 & 1.989 & 2 & \citet{Inada2014} \\
SBS1520+530 & 230.43679 & 52.9135 & 0.761 & 1.855 & 1.6 & 2 & \citet{Chavushyan1997, Auger2008}  \\
SDSSJ1524+4409 & 231.1901 & 44.16377 & 0.320 & 1.210 & 1.67 & 2 & \citet{Oguri2008} \\
J1526-1400 & 231.6891 & -14.0030 & 0.096 & 0.648 & 2.65 & 2 & \citet{Lemon2023} \\
SDSSJ1537+3014 & 234.3935 & 30.24825 & 0.490 & 1.553 & 2.30 & 2 & \citet{More2016} \\
J1548-2914 & 237.1733 & -29.2351 & 0.380 & 1.545 & 1.99 & 2 & \citet{Lemon2023} \\
J1602+4526 & 240.70535 & 45.43528 & 0.43 & 2.16 & 2.70 & 2 & \citet{Lemon2018} \\
SDSSJ1620-1203 & 245.10931 & 12.0613 & 0.398 & 1.158 & 2.77 & 2 & \citet{Kayo2010} \\
FBQ1633+3134 & 248.45408 & 31.56997 & 0.684 & 1.52 & 0.66 & 2 & \citet{Morgan2001} \\
SDSSJ1640-1932 & 250.19029 & 19.54921 & 0.195 & 0.778 & 5.1 & 4 & \citet{Wang2017} \\
SDSSJ1650+4251 & 252.681 & 42.86369 & 0.577 & 1.541 & 1.2 & 2 & \citet{Morgan2003} \\
GraLJ1651-0417 & 252.772 & -4.29027 & 0.591 & 1.451 & 10.1 & 4 & \citet{Stern2021} \\
B1938+666 & 294.60538 & 66.81471 & 0.881 & 2.059 & 0.95 & 4 & \citet{Patnaik1992} \\
WGD2021-4115 & 305.414396 & -41.265865 & 0.335 & 1.390 & 2.75 & 2 & \citet{Agnello2018c} \\
WFJ2033-4723 & 308.4257 & -47.3956 & 0.661 & 1.66 & 2.5 & 4 & \citet{Morgan2004} \\
WGD2038-4008 & 309.511278 & -40.137107 & 0.230 & 0.777 & 2.5 & 4 & \citet{Agnello2018c, Stern2021} \\
B2045+265 & 311.83479 & 26.73367 & 0.8673 & 1.28 & 1.9 & 4 & \citet{Fassnacht1999} \\
J2100-4452 & 315.06191 & -44.86851 & 0.2033 & 0.920 & 1.70 & 4 & \citet{Agnello2019, Spiniello2019} \\
J2106-4944 & 316.5070 & -49.7482 & 0.290 & 1.296 & 2.09 & 2 & \citet{Lemon2023} \\
SDSSJ2146-0047 & 326.6918 & -0.7957 & 0.799 & 2.381 & 1.39 & 2 & \citet{Agnello2015, More2016} \\
HE2149-2745 & 328.0311 & -27.5304 & 0.603 & 2.033 & 1.70 & 2 & \citet{Wisotzki1996, Eigenbrod2006} \\
VDESJ2325-5229 & 351.4217 & -52.4875 & 0.400 & 2.739 & 2.9 & 2 & \citet{Ostrovski2017} \\
    \hline
\end{longtable}
    \small
    \textbf{Note.} The columns are: 

    (1) Lens name; 
    (2) Right Ascension in degrees (J2000); 
    (3) Declination in degrees (J2000); 
    (4) Lens redshift; 
    (5) Source redshift; 
    (6) Einstein radius in arcseconds; 
    (7) Number of images;
    (8) Discovery paper.

\label{lastpage}
\end{document}